# Note sur les temps de service résiduels

## Dans les systèmes type M/G/c

**Thomas Begin[*] - Alexandre Brandwajn[#]**

[*] *UPMC Univ Paris 06, UMR (CNRS) LIP6, Paris, France*
thomas.begin@lip6.fr
[#]*University of California Santa Cruz, Baskin School of Engineering, USA*
alexb@soe.ucsc.edu

RÉSUMÉ. *Lorsque l'analyse exacte de certains modèles paraît difficile, on recherche des solutions approchées. C'est le cas de la file M/G/c pour laquelle les paramètres de performances moyens peuvent être calculés de façon approchée en estimant 1) le minimum des temps restants pour terminer les services des clients en cours de service et 2) la probabilité pour un client arrivant dans la file d'avoir de l'attente. Grâce à une série d'exemples numériques, nous faisons apparaître certaines propriétés méconnues de ces deux quantités. En particulier, pour le minimum des résidus, nous montrons qu'il dépend sensiblement des moments d'ordre supérieur du temps de service, au-delà de la moyenne et du coefficient de variation. Plus généralement, une meilleure connaissance de ces deux quantités, objectif principal de cet article, devrait permettre de mieux saisir quels aspects des approximations des files M/G/c mériteraient d'être repensés afin d'améliorer la qualité de leurs résultats.*

ABSTRACT. *Approximations for the mean performance indices for the M/G/c queue rely on the approximate computation of the probability that an arriving request has to wait for service and of the minimum of residual service times if all servers are found busy. Using numerical examples, we investigate properties of these two quantities. In particular, we show that the minimum of residual service times depends on higher order properties, beyond the first two moments, of the service time distribution. Improved knowledge of the properties of the two quantities studied in this paper provides insight into avenues for improving the accuracy of approximations for the M/G/c queue.*

MOTS-CLÉS : *Files multiserveurs, Temps résiduels, Approximation, Moments d'ordre supérieur.*

# 1. Introduction

Les phénomènes de congestion présents dans de nombreux protocoles, systèmes et réseaux informatiques peuvent être représentés par une ou plusieurs files d'attente [HEI 84, ALL 90]. L'élaboration de ces modèles et leur analyse ont permis de mieux concevoir, dimensionner et calibrer les systèmes informatiques actuels.

Pour des files telles que les M/G/1, M/G/c et G/G/c, le calcul de certains paramètres de performance (par exemple, le temps d'attente moyen ou le nombre moyen de clients en attente) devient simple si l'on est capable d'évaluer *les temps de service résiduels* des clients en cours de service. Les temps de service résiduels sont définis comme les temps restants pour terminer les services des clients en cours de service au moment où un nouveau client entre dans la file. En général, dans le cas d'une file monoserveur, on s'intéresse à la valeur moyenne de ces temps résiduels, $t_r$, appelée le *résidu* de service, et dans le cas des files multiserveurs, on s'intéresse à la valeur moyenne du plus petit des $t_r$ pour l'ensemble des clients en cours de service, c'est-à-dire au *minimum des résidus* $\min(t_r)$.

Le calcul de $t_r$ peut être simple comme c'est le cas pour les files M/M/1 [KLE 75] puisque la propriété sans-mémoire de la distribution exponentielle implique qu'à tout instant le temps moyen qu'il reste au serveur pour terminer son service est égal au temps de service moyen. Dans le cas des files multiserveurs M/M/c [KLE 75], le calcul du minimum des $t_r$ reste simple puisque, ici aussi, la propriété sans-mémoire du temps de service implique qu'à tout instant, le minimum des temps résiduels des $k$ clients en cours de service est identique à celui de $k$ clients qui viendraient de démarrer leur service. Dans le cas d'une file M/G/1, dont le temps de service a une moyenne $m$ et un coefficient de variation $cv$, le temps de service résiduel d'un client est en moyenne égal à $t_r = \frac{m}{2}(1 + cv^2)$ [KLE 75]. Cette expression très compacte du temps de service résiduel permet d'obtenir des formules simples pour les paramètres de performance moyens d'une file M/G/1.

Toutefois, pour la majorité des files monoserveurs où le processus d'arrivée est non poissonien, ainsi que pour l'ensemble des files multiserveurs dès lors que les temps de service des serveurs ne sont pas distribués exponentiellement, il n'existe aucune expression exacte simple du résidu, ou bien du minimum des résidus lorsqu'il y a plusieurs serveurs. Aussi des chercheurs ont proposé des méthodes pour estimer ces quantités. Ces estimations, dont le domaine de validité peut être restreint, ont permis à leurs auteurs d'obtenir des approximations pour les paramètres de performance moyens de la file considérée [MIY 86, NOZ 78, LAZ 86, EAG 00]. Dans le cas des réseaux fermés, lorsque les hypothèses d'application de l'algorithme MVA (« Mean Value Analysis ») [REI 79, REI 81] ne sont pas rassemblées, il existe des méthodes pour obtenir une solution approchée. Certaines de ces méthodes [HAL 00, LAZ 86, EAG 00] font appel à une estimation des temps résiduels de service qui est calquée sur celle présentée ci-dessus pour la file M/G/1. Pour les files multiserveurs, plusieurs auteurs [MIY 86, HOK 78, HOO 86, MOR 80, NOZ 78] ont proposé des solutions approchées pour le calcul du minimum des résidus en adaptant la formule exacte pour la file M/G/1 au cas multiserveur. Ces études ont permis d'obtenir des approximations pour les paramètres de performance moyens des files M/G/c ou G/G/c. Toutefois ces approximations ont des domaines d'applicabilité restreints car les estimations du minimum des résidus peuvent dans certains cas s'écarter sensiblement de sa valeur réelle.

Dans cet article, nous nous concentrons particulièrement sur le comportement du minimum des temps de service résiduels $t_r$ dans une file multiserveur à arrivées poissoniennes. Nous obtenons les valeurs de ces quantités grâce à des simulations que nous avons elles-mêmes validées par rapport à une solution semi-numérique exacte permettant le calcul des probabilités des états stationnaires [BRA 07]. Dans la Section 2, nous décrivons la file considérée et nous présentons une technique « classique » pour obtenir une approximation des performances moyennes d'une telle file. Ce type de technique repose sur l'estimation de plusieurs quantités, et notamment sur celle du minimum des résidus. La Section 3 contient les résultats numériques et met en lumière certaines dépendances fortes et méconnues entre le minimum des résidus et les paramètres de la file. En particulier, contrairement à une idée largement admise et reprise en tant qu'hypothèse par les approximations, nous montrons que le minimum des résidus ne dépend pas uniquement de la moyenne et du coefficient de variation du temps de service mais que les moments d'ordre supérieur du temps de service constituent un facteur important dans l'évaluation de cette quantité. Ces dépendances qui semblent méconnues contribuent à la difficulté d'établir correctement le domaine d'applicabilité des approximations pour la file M/G/c. La Section 4 conclut cet article.

## 2. Système considéré

Tout au long de cet article, nous considérons une file multiserveur à capacité infinie avec discipline de service FIFO. Les arrivées des clients dans cette file sont engendrées par un processus de Poisson de taux $\lambda$. La file est composée de $c$ serveurs identiques dont les temps de service sont en moyenne égaux à $m$. Nous représentons la loi de service de chacun des serveurs par une distribution de Cox à deux étages [COX 61] comme illustré sur la Figure 1. Les taux de service du premier et du second étage de cette distribution sont respectivement donnés par $\mu_1$ et $\mu_2$. $\hat{q}_1$ représente la probabilité pour qu'un serveur poursuive le service d'un client après la conclusion du premier étage. Réciproquement, la probabilité pour qu'un client, une fois le premier étage terminé, quitte la file est donné par $q_1 = 1 - \hat{q}_1$. Le modèle décrit est donc un cas particulier de la file M/G/c pour laquelle le coefficient de variation du temps de service des serveurs est noté $cv$.

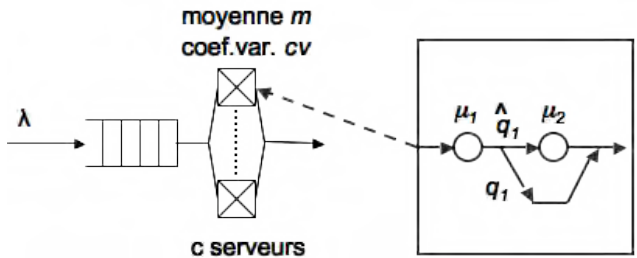

**Figure 1.** *Modèle considéré*

Nous nous intéressons à l'évaluation des paramètres de performance moyens d'une telle file multiserveur en régime stationnaire. L'utilisation moyenne des serveurs est égale à $\rho = \lambda m/c$ ($\rho$ varie entre 0 et 1). En équilibre, le débit moyen en sortie d'une file sans perte en entrée est connu et toujours égal au débit moyen en entrée, i.e. $\lambda$. Le nombre moyen stationnaire de clients dans la file $E(Q)$ peut être exprimé en fonction de $\rho$ et du nombre moyen de clients en attente de libération d'un des serveurs $E(Q_w)$. La loi de Little [KLE 75] permet de remplacer $E(Q_w)$ par le produit du temps qu'un client attend en moyenne avant de démarrer son service, $E(W)$, et du débit moyen $\lambda$. Ainsi, on obtient la relation suivante pour $E(Q)$ :

$E(Q) = c\rho + E(W)\lambda$

Si l'on sait évaluer le temps moyen d'attente avant service d'un client $E(W)$, alors la dérivation des autres paramètres de performance moyens pour une file multiserveur à capacité infinie devient immédiate. Il paraît difficile dans le cas général d'obtenir une formule simple pour $E(W)$. En revanche, certains travaux ont proposé des approximations pour $E(W)$. Soit $\Pi_w$ la probabilité qu'un client arrivant dans la file trouve tous les serveurs occupés. Rappelons que $\min(t_r)$ représente la valeur moyenne du minimum des résidus des temps de service des clients en cours de service. $E(W)$ peut s'estimer comme :

$E(W) \approx \Pi_w \min(t_r) + mE(Q_w)/c$

En appliquant à nouveau la formule de Little sur $E(Q_w)$, on obtient :

$$E(W) \approx \frac{\Pi_w \min(t_r)}{1-\rho} \qquad (1)$$

Ainsi, si l'on connaît les deux quantités $\Pi_w$ et $\min(t_r)$, l'evaluation de $E(W)$ à partir de (1) ainsi que celle des paramètres de performance moyens d'une file multiserveur à capacité infinie est immédiat. Malheureusement, dans le cas général, il est difficile d'obtenir une expression exacte de ces deux quantités. C'est pourquoi de nombreux travaux en ont proposé des estimations afin d'obtenir des approximations des paramètres de performance moyens pour des files telles que la M/G/c et la G/G/c [MIY 86, HOK 78, HOO 86, MOR 86, NOZ 78].

Hormis l'approximation inhérente à la formule (1), la qualité de ces approximations est directement liée à la précision des estimations de $\Pi_w$ et $\min(t_r)$. Or, il semblerait que peu de travaux se soient intéressés à évaluer cette précision et à établir les zones d'applicabilité de ces estimations.

A l'aide d'un ensemble d'exemples numériques présentés dans la section suivante, nous mettons en lumière certaines limites de ces approximations, dues notamment à des dépendances fortes et non prises en compte dans l'évaluation approximative de $\min(t_r)$. Pour ce faire, nous utilisons une solution semi-numérique exacte des probabilités des états stationnaires d'une file M/G/c du type considéré [BRA 07], pour obtenir la valeur exacte de la quantité $\Pi_w$. Concernant le minimum des residus $\min(t_r)$, nous obtenons sa valeur exacte grâce à des simulations à événements discrets dont nous avons vérifié l'intégrité par rapport à la solution semi-numérique exacte précédemment citée.

## 3. Résultats numériques
### 3.1 *Minimum des résidus*

Dans tous les exemples de cette section, les temps de service moyens des serveurs sont exprimés dans une unité temporelle quelconque et égaux à 1 (on rappelle que l'on désigne le moment d'ordre 1 du temps de service par $m$). Pour chaque étude, nous indiquons le coefficient de variation, le coefficient de dissymétrie (skewness), le coefficient d'aplatissement (kurtosis) du temps de service (quantités associées respectivement aux moments d'ordre 2, 3 et 4), ainsi que la description complète de la distribution type Cox-2 le représentant. Pour chaque figure obtenue par simulation, nous avons représenté à l'aide d'un trait fin les intervalles de confiance estimés à 95%.

3.1.1 *Dépendance vis-à-vis de la distribution du temps de service*

Nous commençons par évaluer les valeurs du minimum des résidus d'une file M/G/c avec 4 serveurs pour une utilisation moyenne de ses serveurs égale à 0.5, i.e. $\rho$=0.5. La Figure 2 représente l'évolution de $\min(t_r)$ pour des valeurs croissantes du coefficient de variation $cv$. Les distributions type Cox-2 utilisées dans cet exemple sont décrites dans le Tableau 1. Notons que le taux $\mu_1$ du premier étage de ces distributions est toujours de 1000. Dans la suite de l'article, on désignera cet ensemble de distributions par Dist. I.. Pour les distributions du Tableau 1, les résultats de la Figure 2 semblent indiquer que $\min(t_r)$ augmente lorsque $cv$ augmente, comme c'est également le cas pour la file M/G/1.

| $cv$ | $\mu_1$ | $\mu_2$ | $q_1$ | Skewness | Kurtosis |
|---|---|---|---|---|---|
| 2 | 1000 | 0.400 | 0.601 | 3.07 | 12.77 |
| 4 | 1000 | 0.118 | 0.883 | 6.01 | 48.28 |
| 6 | 1000 | 0.054 | 0.946 | 9.01 | 108.30 |
| 8 | 1000 | 0.031 | 0.969 | 12.01 | 192.43 |
| 10 | 1000 | 0.020 | 0.980 | 15.02 | 300.63 |

**Tableau 1.** *Distributions I*

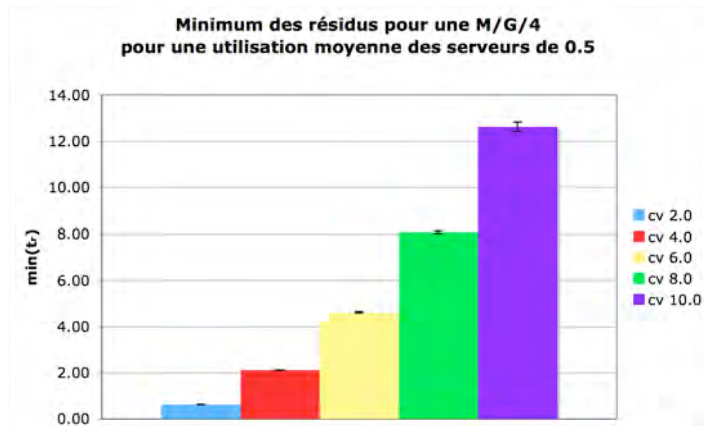

**Figure 2.** *Influence du coefficient de variation pour Dist. I*

Pour l'étude suivante, nous considérons également une file M/G/4 avec $\rho$=0.5. Nous évaluons $\min(t_r)$ avec les mêmes valeurs de $m$ et de $cv$ que dans l'exemple précédent, mais les distributions des temps de service ne sont pas les mêmes. La description de ces distributions, que l'on désigne par Dist. II, se trouve dans le Tableau 2. Rappelons que les temps de service considérés dans cet article varient par leur distribution mais que leur moyenne reste toujours égale à 1.0. Contrairement à l'exemple précédent, les $\min(t_r)$ pour cet ensemble de distributions sont petits et restent quasi-constants lorsque le $cv$ du temps de service change comme le montre la Figure 3. La Figure 4 rassemble les résultats des deux exemples précédents sur une seule figure et montre clairement que les moments d'ordre 1 et 2 du temps de service ne suffisent pas pour estimer correctement la valeur de $\min(t_r)$. Par exemple, pour un $cv$ de 4, $\min(t_r)$ peut passer de 0.27 à 2.12 selon que la distribution utilisée pour représenter le temps de service est choisie parmi les Dist. I ou II. On conclut de ces exemples que, à l'inverse d'une file M/G/1, dans les files M/G/c les moments d'ordre supérieur à 2 du temps de service interviennent dans le calcul du minimum des résidus des temps de service en cours que « voit » un client qui rejoint la file.

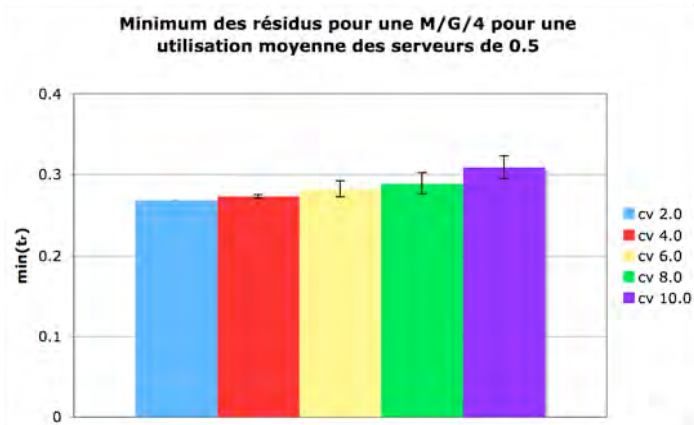

**Figure 3.** *Influence du coefficient de variation pour Dist. II*

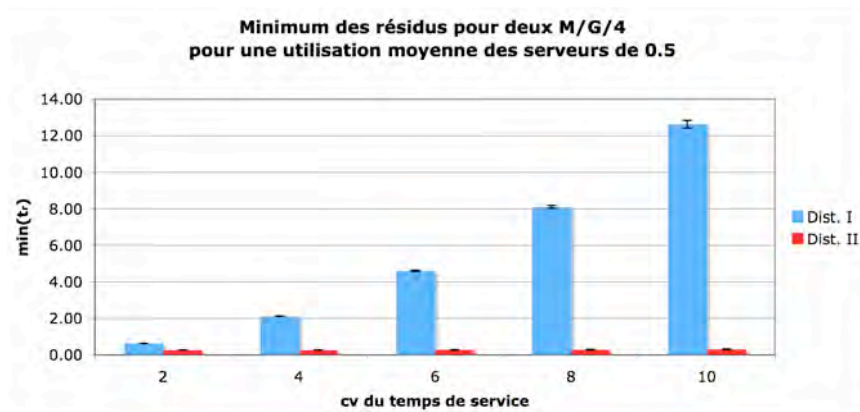

**Figure 4.** *Influence des moments d'ordre supérieur*

| $cv$ | $\mu_1$ | $\mu_2$ | $q_1$ | Skewness | Kurtosis |
|---|---|---|---|---|---|
| 2 | 1.11 | 0.063 | 0.9938 | 19.26 | 608.91 |
| 4 | 1.11 | 0.013 | 0.9987 | 54.10 | 4107.30 |
| 6 | 1.11 | 0.006 | 0.9994 | 86.00 | 10087.28 |
| 8 | 1.11 | 0.003 | 0.9997 | 116.98 | 18480.19 |
| 10 | 1.11 | 0.002 | 0.9998 | 147.58 | 29276.89 |

**Tableau 2.** *Distributions II*

3.1.2 *Dépendance vis-à-vis de l'utilisation moyenne des serveurs*

A présent, nous vérifions que le minimum des résidus dans une file M/G/c est indépendant de l'intensité de la charge $\rho$ en entrée, comme c'est le cas pour une file monoserveur M/G/1. Pour cela, nous considérons une file M/G/4 à plusieurs niveaux d'utilisation de ses serveurs. La Figure 4 représente $\min(t_r)$ pour les deux distributions du temps de service ayant un $cv$ égal à 4 dans Dist. I et II. Sur cet exemple, et plus généralement sur tous les exemples que nous avons effectués pour des files M/G/c avec des valeurs différentes pour c et $cv$, nous avons constaté que la valeur de $\min(t_r)$ est indépendante du débit moyen en entrée, ou de manière équivalente, qu'elle ne dépend pas de l'utilisation moyenne des serveurs, c'est-à-dire de $\rho$.

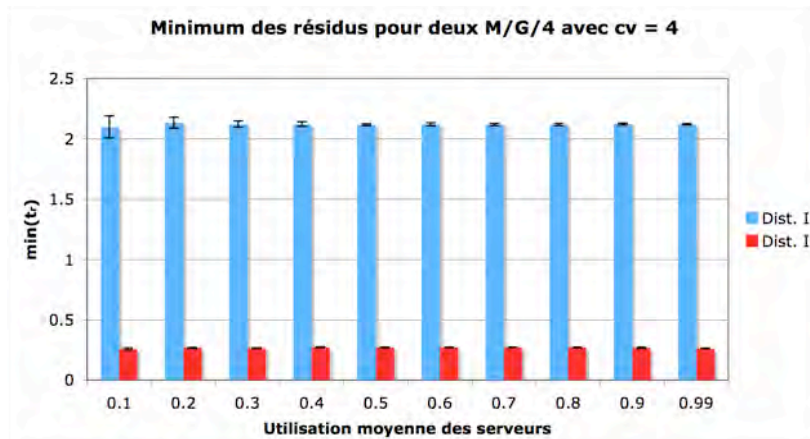

**Figure 5.** *Minimum des résidus pour plusieurs niveaux de charge en entrée*

3.1.3 *Dépendance vis-à-vis du nombre de serveurs*

A présent, nous nous intéressons à l'évolution de $\min(t_r)$ en fonction du nombre de serveurs dans la file M/G/c. Nous considérons trois distributions différentes du temps de service, mais qui ont toutes la même moyenne et le même coefficient de variation $cv$. Elles correspondent aux trois distributions ayant $cv$=4 dans Dist. I, II et III. Notons que dans nos calculs nous avons fixé $\rho$ à 0.5 mais, étant donné les résultats précédents, la valeur de $\rho$ n'importe pas. La Figure 6 représente les résultats obtenus pour un nombre croissant de serveurs allant de 2 à 10. Nous observons assez logiquement que plus le nombre de serveurs est grand, plus le minimum des résidus des services en cours est petit. Plus intéressant, la figure montre que la vitesse de cette décroissance avec c varie selon la distribution choisie pour le temps de service. Par exemple, le minimum des résidus $\min(t_r)$ est réduit d'un facteur d'environ 2.0 lorsque c passe de 2 à 4 pour la distribution choisie dans Dist. I, tandis que ce facteur passe à 3.9 pour celle choisie dans Dist. III. Plus généralement, nos résultats montrent que cette décroissance n'est pas linéaire, sauf dans certains cas comme ici pour Dist. I. Pour Dist. II et III, la décroissance est plus rapide et elle s'intensifie lorsque c augmente. Ainsi, la décroissance de $\min(t_r)$ avec c dépend des propriétés d'ordre supérieur à 2 de la distribution des temps de service.

| $cv$ | $\mu_1$ | $\mu_2$ | $q_1$ | Skewness | Kurtosis |
|---|---|---|---|---|---|
| 2 | 2.5 | 0.286 | 0.829 | 4.64 | 29.76 |
| 4 | 2.5 | 0.074 | 0.956 | 9.80 | 129.44 |
| 6 | 2.5 | 0.033 | 0.980 | 14.87 | 296.05 |
| 8 | 2.5 | 0.019 | 0.989 | 19.90 | 529.36 |
| 10 | 2.5 | 0.012 | 0.993 | 24.92 | 829.35 |

**Tableau 3.** *Distributions III*

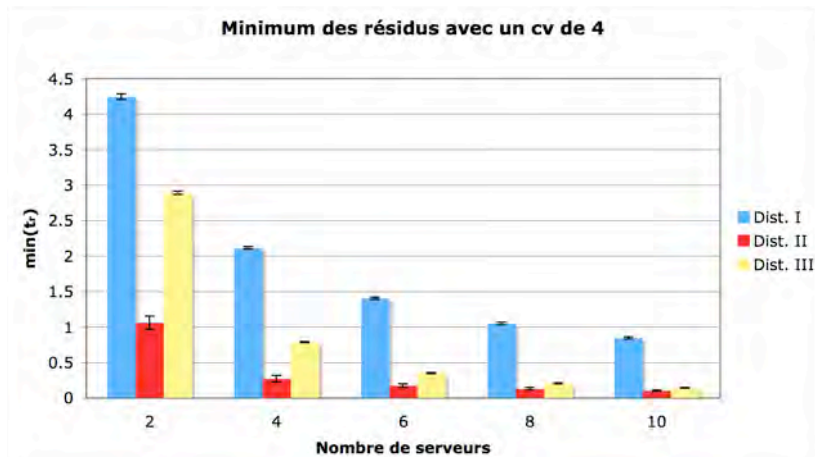

**Figure 6**. *Influence du nombre de serveurs sur le minimum des résidus*

3.1.4 *Erreur liée à certaines approximations*

Jusqu'ici nous avons montré que le minimum des résidus dans une file M/G/c ne dépend pas uniquement des deux premiers moments *m* et *cv* du temps de service. Nos résultats montrent aussi que $\min(t_r)$ a tendance à diminuer lorsque c augmente et que la vitesse de cette diminution varie selon les caractéristiques d'ordre supérieur des distributions du temps de service. Enfin, nos expérimentations corroborent l'idée que, comme on pouvait s'y attendre, $\min(t_r)$ ne dépend pas de l'utilisation moyenne des serveurs d'une M/G/c.

Un certain nombre de solutions approchées pour évaluer les paramètres de performances moyens d'une M/G/c reposent sur des estimations de $\min(t_r)$ et $\Pi_w$ d'après la formule (1). Pour $\min(t_r)$, les résultats présentés ci-dessus s'opposent à certaines hypothèses faites par ces approximations. D'une part, la plupart de ces approximations [MIY 86, NOZ 78, LAZ 86, EAG 00] estime la quantité $\min(t_r)$ en prenant en compte uniquement les deux premiers moments du temps de service. Notons que les imprécisions de ces approximations, signalées dans quelques articles [MA 95, BRA 07], s'expliquent peut-être par l'erreur commise initialement dans l'estimation de $\min(t_r)$. D'autre part, certaines approximations estiment la valeur de $\min(t_r)$ pour une file M/G/c [MOR 80, HOK 78] en divisant par c la valeur du temps résiduel trouvée pour une M/G/1 de $m$ et $cv$ identiques. Ceci s'apparente à approcher le minimum de trois variables aléatoires indépendantes par une seule qui serait trois fois plus rapide. Or, si nous avons observé que $\min(t_r)$ a tendance à diminuer lorsque c croît, nous avons également montré que cette décroissance n'est, en général, pas linéaire. On s'attend donc à ce que les approximations qui ne tiennent compte que de $m$ et de $cv$ ainsi que celles qui supposent une relation simple et linéaire entre $\min(t_r)$ et c puissent s'éloigner sensiblement de la valeur exacte de $\min(t_r)$ pour certaines distributions.

La formule la plus simple pour approcher $\min(t_r)$, calquée sur la formule exacte pour estimer $t_r$ dans le cas d'une file M/G/1, et utilisée par certaines approximations [MOR 80, HOK 78] est :

$$\min(t_r) \approx m \frac{(1+cv^2)}{2c} \qquad (2)$$

Nous étudions maintenant l'erreur liée à cette approximation de $\min(t_r)$ pour toutes les distributions de Dist. I, II et III. Les Figures 7 et 8 représentent respectivement les écarts relatifs obtenus dans le cas d'une file M/G/2 et d'une file M/G/4. Il est intéressant de noter que dans le cas de Dist. I, l'approximation délivre de bons résultats avec des écarts relatifs inférieurs à 1%, tandis que pour Dist. II et III, les écarts sont importants. Par exemple, si on considère la distribution avec un $cv$ de 2 de Dist. II, l'écart atteint environ 100%. Nous observons également que l'erreur semble être plus importante pour les grands $cv$.

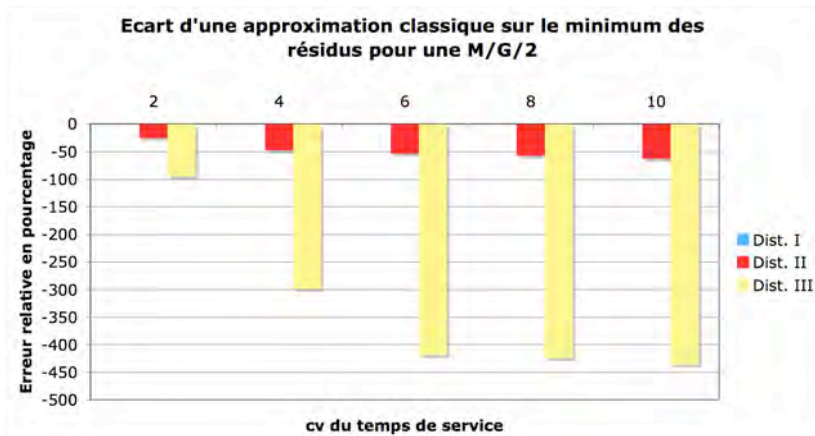

**Figure 7**. Erreur relative de l'approximation du minimum des résidus

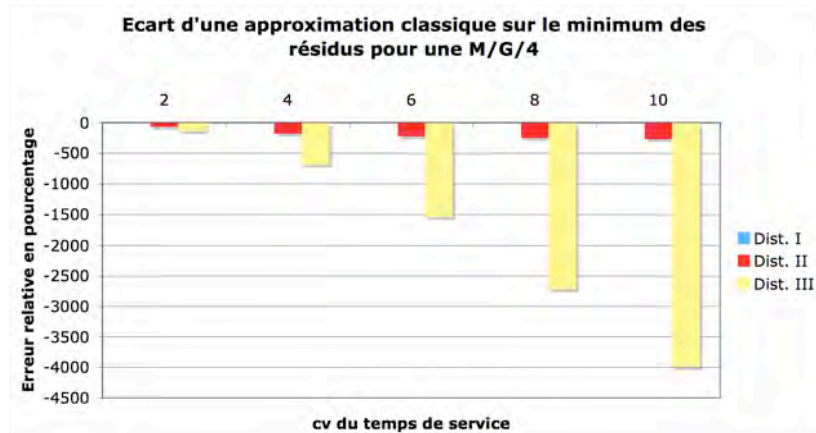

**Figure 8**. *Erreur relative de l'approximation du minimum des résidus*

**3.2** *Probabilité d'attente*

A présent nous nous intéressons à l'autre quantité engagée dans l'évaluation approchée des performances moyennes d'une file M/G/c d'après la formule (1). Il s'agit de $\Pi_w$ qui désigne la probabilité qu'un client qui arrive dans la file ait de l'attente. Son calcul exact apparaissant compliqué pour une file M/G/c, l'hypothèse de pouvoir remplacer $\Pi_w$ par la probabilité d'attente dans une M/M/c de même taux est attractive et a été souvent employée [TIJ 81, MIY 86, MOR 80]. Nous savons que cette probabilité, en tant que somme des probabilités des états stationnaires de 0 à c-1, dépend du coefficient de variation et également des moments d'ordre supérieur du temps de service [BRA 07]. Toutefois, nos expériences ont montré que l'écart entre les probabilités d'attente dans les deux files reste modéré (ordinairement compris entre 5 et 10% et pouvant aller jusqu'à 15%) pour un nombre de serveurs allant de 2 à 10, pour des utilisations moyennes des serveurs prises entre 0.1 et 0.99 et des distributions du temps de service choisies parmi Dist. I, II et III. La Figure 9 représente l'erreur liée à l'approximation pour une file avec 4 serveurs et des temps de service choisis dans Dist. III. Nous avons observé que ces écarts ont tendance à s'accroître lorsque *cv* est grand, et plus intéressant peut-être, nous avons constaté que les erreurs les plus grandes ont lieu pour des utilisations des serveurs de l'ordre de 0.5.

En conclusion de cette section, nous pensons que l'approximation de la valeur de $\Pi_w$ dans une M/G/c par sa valeur dans une M/M/c convient aux exigences d'une solution approchée qui s'appuie sur la formule (1) pour obtenir les paramètres de performances moyens d'une M/G/c. Concernant l'estimation de $\min(t_r)$, l'omission des moments d'ordre supérieur du temps de service ainsi que la prise en compte parfois trop approximative du nombre de serveurs, peut conduire à des erreurs importantes qui se répercutent dans le calcul des paramètres de performances moyens de la file M/G/c.

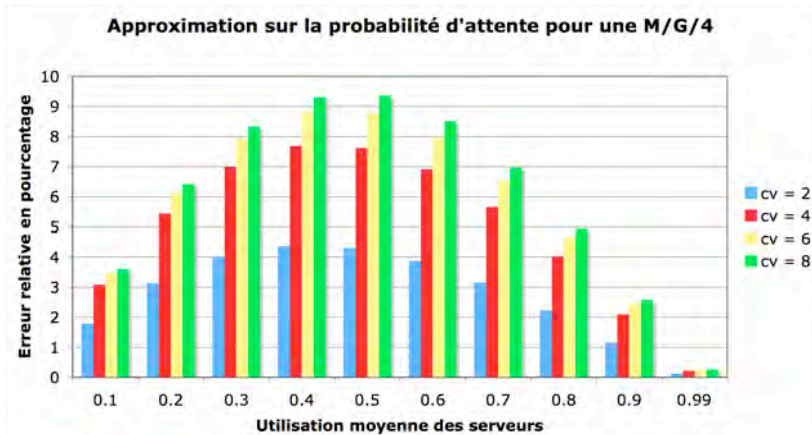

**Figure 9**. *Erreur relative de l'approximation sur la probabilité $\Pi_w$ pour Dist. III*

## 4. Conclusions

Dans cet article, nous nous sommes intéressés à deux quantités : le minimum des résidus et la probabilité d'attente d'un client, qui doivent être estimées pour pouvoir appliquer certaines approximations des paramètres de performances moyens d'une file M/G/c [MOR 80, MIY 86, HOO 86, HOK 78, NOZ 78, TIJ 81].

Nous avons montré que le minimum des résidus ne dépend pas uniquement des deux premiers moments du temps de service comme c'est le cas pour une file M/G/1. Des valeurs très différentes du minimum des résidus sont possibles pour des distributions du temps de service ayant la même moyenne et le même coefficient de variation. Toujours pour le minimum des résidus, nous avons montré que son évolution avec le nombre de serveurs dans la file dépend de la distribution du temps de service. Par conséquent, il n'est pas surprenant que les approximations du minimum des résidus qui dérogent à ces propriétés puissent aboutir à des erreurs importantes, y compris pour des valeurs modérées du coefficient de variation de la distribution du temps de service (par exemple, 300% d'erreur relative pour un *cv* de 4 dans une M/G/2). Par ailleurs, nos résultats ont confirmé que la valeur du minimum des résidus est indépendante de l'utilisation moyenne des serveurs, comme c'est aussi le cas pour la file M/G/1.

Concernant la probabilité d'attente d'un client qui arrive dans une M/G/c, nos résultats permettent de vérifier que la solution courante qui consiste à remplacer cette probabilité par celle obtenue pour une M/M/c de même taux conduit à des écarts modérés, y compris pour des *cv* importants (environ 5 à 10% d'erreur relative pour des *cv* allants jusqu'à 8).

Pour conclure, les résultats de nos travaux semblent indiquer que l'amélioration des approximations, pour la file M/G/c, considérées dans cet article impliquent principalement une meilleure estimation du minimum des résidus. Une suite logique à ces travaux consisterait à étendre cette étude à d'autres types de distributions du temps de service et à rechercher quels sont les moments du temps de service dont il faut tenir compte pour obtenir une bonne estimation - voire la valeur exacte - de cette quantité.

## 5. Bibliographie

[ALL 90] ALLEN A. O., *Probability, statistics and queuing theory with computer science applications,* Vol. 2, San Diego: AcademicPress, 1990.

[BRA 07] BRANDWAJN, A., BEGIN, T., « A Novel Approach to G/C2/c-type queues », Rapport Technique, 2007, University of California Santa Cruz.

[COX 61] COX, D. R., SMITH, W. L. *Queues.* John Wiley, New York, 1961.

[EAG 00] EAGER, D. L., SORIN, D. J., VERNON, M. K. « AMVA techniques for high service time variability », *SIGMETRICS, Perform. Eval. Rev.* Vol. 28, No. 1, 2000, p. 217-228.

[HAL 00] HALACHMI, I., ADAN, I. J. B. F., VAN DER WAL, J., HEESTERBEEK, J. A. P., VAN BEEK, P, « The design of robotic dairy barns using closed queueing networks », *European Journal of Operational Research*, Elsevier, Vol. 124, No. 3, 2000, p. 437-446.

[HEI 84] HEIDELBERGER P., LAVENBERG, S.S., « Computer performance evaluation methodology », *IEEE Trans. Computers*, Vol. 33, 1984, p. 1195-1220.

[HOK 78] HOKSTAD, P., « Approximations for the M/G/m Queue » *Operations Research*, Vol. 26, No. 3. (May - Jun., 1978), p. 510-523.

[HOO 86] VAN HOORN M. H., SEELEN, L. P., « Approximations for the GI/G/c Queue », *Journal of Applied Probability*, Vol. 23, No. 2, 1986, p. 484-494.

[KLE 75] KLEINROCK L., *Queueing Systems Volume 1 : Theory*, Wiley, 1975.

[LAZ 86] LAZOWSKA E. D., ZAHORJAN J., SEVCIK, K. C., « Computer System Performance Evaluation Using Queueing Network Models », *Annual Review of Computer Science*, Vol. 1, 1986.

[MA 95] MA B. N. W., AND MARK J. W., « Approximation of the Mean Queue Length of an M/G/c Queueing System », *Operations Research*, Vol. 43, No. 1, Special Issue on Telecommunications Systems: Modeling, Analysis and Design, 1995, p. 158-165.

[MIY 86] MIYAZAWA, M., « Approximation of the Queue-Length Distribution of an M/GI/s Queue by the Basic Equations », *Journal of Applied Probability*, Vol. 23, No. 2, 1986, p. 443-458.

[MOR 80] MORI M., « Relations between Queue-Size and Waiting-Time Distributions », *Journal of Applied Probability*, Vol. 17, No. 3, 1980, p. 822-830.

[NOZ 78] NOZAKI, S. A., ROSS, S. M., « Approximations in Finite-Capacity Multi-Server Queues with Poisson Arrivals, » *Journal of Applied Probability*, Vol. 15, No. 4, 1978, p. 826-834.

[REI 79] REISER, M., « Mean Value Analysis for Queueing Networks - A New Look at an Old Problem », P*roceedings of the 3rd International Symposium on Modelling and Performance Evaluation of Computer Systems*, 1979, p. 63-77.

[REI 81] REISER, M., « Mean-Value Analysis and Convolution Method for Queue-Dependent Servers in Closed Queueing Networks », *Performance Evaluation,* Vol. 1, 1981, p. 7-18.

[TIJ 81] TIJMS, H. C., VAN HOORN, M. H., FEDERGRUEN, A., « Approximations for the Steady-State Probabilities in the M/G/c Queue », *Advances in Applied Probability*, Vol. 13, No. 1, 1981, p. 186-206.